\documentclass[conference]{IEEEtran}
\IEEEoverridecommandlockouts
\usepackage{cite}
\usepackage{amsmath,amssymb,amsfonts}
\usepackage{algorithmic}
\usepackage{graphicx}
\usepackage{textcomp}
\usepackage{multirow}
\usepackage{booktabs}
\usepackage[inline]{enumitem}
\usepackage{xcolor}
\usepackage{cleveref}
\def\BibTeX{{\rm B\kern-.05em{\sc i\kern-.025em b}\kern-.08em
    T\kern-.1667em\lower.7ex\hbox{E}\kern-.125emX}}
\begin{document}

\title{Architectural Vision for Quantum Computing in the Edge-Cloud Continuum
}

\makeatletter
\newcommand{\linebreakand}{%
  \end{@IEEEauthorhalign}
  \hfill\mbox{}\par
  \mbox{}\hfill\begin{@IEEEauthorhalign}
}
\makeatother 


\thispagestyle{plain}
\pagestyle{plain}
\author{
    Alireza Furutanpey$^{\dagger*}$,
    Johanna Barzen$^\ddagger$,
    Marvin Bechtold$^\ddagger$, 
    Schahram Dustdar$^\dagger$, \\
    Frank Leymann$^\ddagger$,
    Philipp Raith$^\dagger$,
    Felix Truger$^\ddagger$
    \\
    \small{$^\dagger$Distributed Systems Group, TU Vienna} \texttt{first.last@dsg.tuwien.ac.at} 
    \\
    \small{$^\ddagger$Institute of Architecture of Application Systems, University of Stuttgart} \texttt{first.last@iaas.uni-stuttgart.de}
    \\
    \small{$^{*}$\textbf{Corresponding Author}}
    \
}
\maketitle

\maketitle

\begin{abstract}
Quantum processing units (QPUs) are currently exclusively available from cloud vendors. However, with recent advancements, hosting QPUs is soon possible everywhere. Existing work has yet to draw from research in edge computing to explore systems exploiting mobile QPUs, or how hybrid applications can benefit from distributed heterogeneous resources. Hence, this work presents an architecture for Quantum Computing in the edge-cloud continuum. We discuss the necessity, challenges, and solution approaches for extending existing work on classical edge computing to integrate QPUs. We describe how warm-starting allows defining workflows that exploit the hierarchical resources spread across the continuum. Then, we introduce a distributed inference engine with hybrid classical-quantum neural networks (QNNs) to aid system designers in accommodating applications with complex requirements that incur the highest degree of heterogeneity. We propose solutions focusing on classical layer partitioning and quantum circuit cutting to demonstrate the potential of utilizing classical and quantum computation across the continuum. To evaluate the importance and feasibility of our vision, we provide a proof of concept that exemplifies how extending a classical partition method to integrate quantum circuits can improve the solution quality. Specifically, we implement a split neural network with optional hybrid QNN predictors. Our results show that extending classical methods with QNNs is viable and promising for future work. 
\end{abstract}


\section{Introduction} \label{sec:intro}
\begin{figure}[ht]
\centering
\includegraphics[width=\columnwidth]{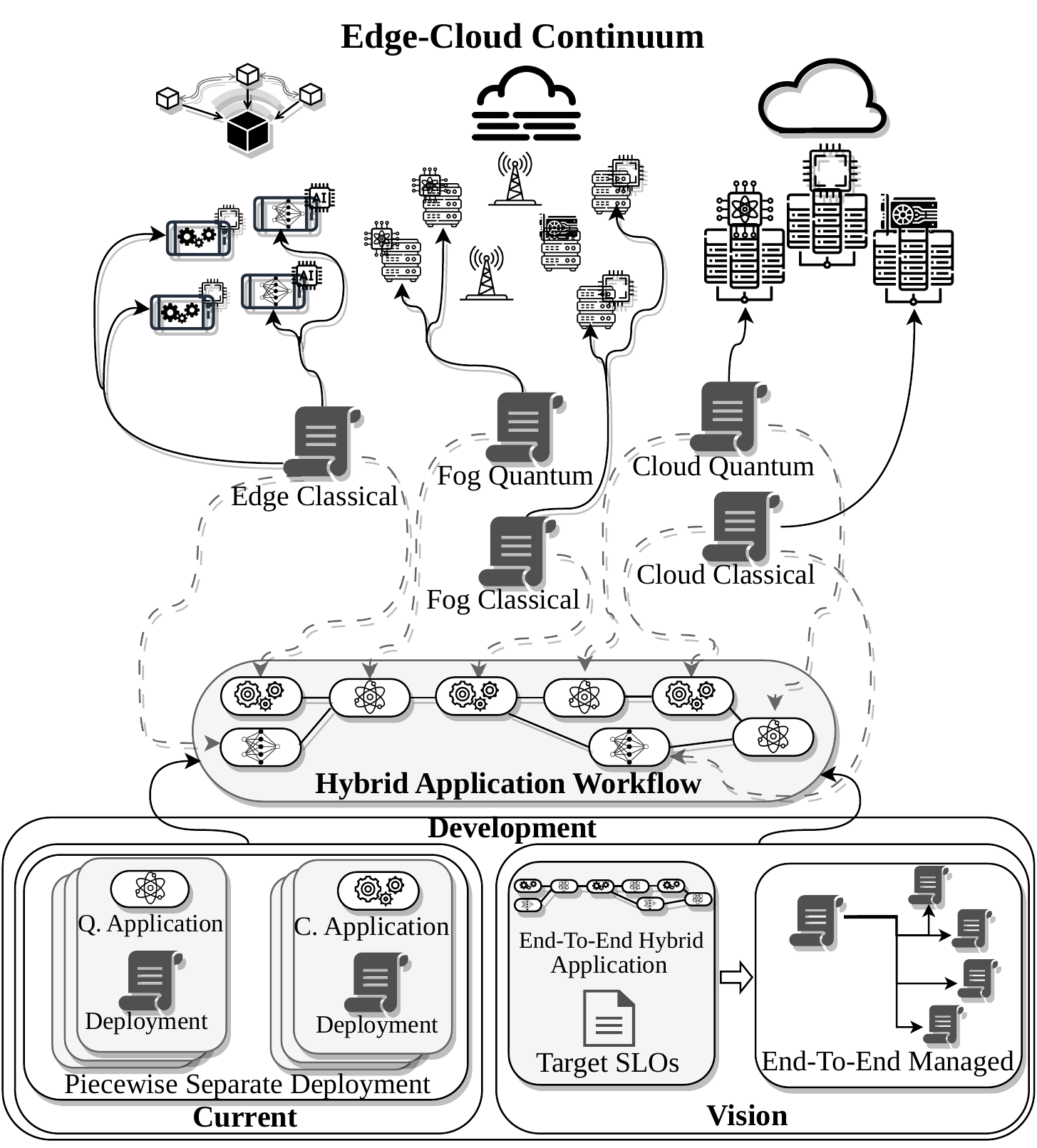}
\caption{A Distributed Classical-Quantum Hybrid Platform}
\label{fig:vision}
\end{figure}
Noisy intermediate-scale quantum (NISQ) computers are error-prone, contain only a limited number of qubits, and impose restrictions on the depth of successfully executable circuits~\cite{nisqbitter}. Yet, algorithms tailored towards NISQ devices started to demonstrate the viability of quantum computers in various fields, ranging from molecule simulation~\cite{moleculeapp} to machine learning~\cite{qmlapp} and optimization problems~\cite{qaoaapp}. Evidently, to advance research and development into practical applications of quantum algorithms, increasing the accessibility of quantum computers by introducing adequate abstractions is effective. Nevertheless, providing researchers and practitioners access to QPUs is challenging. Owing to the limited availability, complexity, and cost of Quantum Processing Units (QPUs), quantum computation for the masses may currently only be viable through cloud services that can hide the low-level machinery behind a convenient interface. 
However, while cloud providers can decrease the complexity and cost, we cannot exclusively rely on the efforts of hardware manufacturers to increase the accessibility of resources by simplifying the production and installation of QPUs. Instead, it is essential to draw from our experiences in classical computing on the ramifications of predominantly relying on centralized cloud platforms. Besides the privacy-related risks of entrusting third-party providers with sensitive data, the cloud computing paradigm bears numerous downsides, such as vendor lock-in and data centers posing a single point of failure vulnerable to outages. Additionally, a narrow cloud-centric view is inefficient since it leaves valuable resources close to the client (e.g., Edge, Fog) idle by indiscriminately offloading tasks to a remote server. 

The edge-cloud continuum addresses the limitations of cloud computing by aggregating resources in a hierarchical distributed network ranging from constrained edge devices to cloud data centers. Unfortunately, after decades of relying on centralized architectures, the transition is slow, with semi- or fully decentralized platforms still needing to emerge~\cite{nastic2022serverless}. Therefore, to avoid repeating the mistakes of classical computing, it is crucial to aid researchers in implementing quantum applications on distributed platforms before centralized approaches solidify.  

Notably, deploying edge applications on mobile quantum devices is on the horizon with the recent advancements of diamond-based QPUs~\cite{saxonq} that allow quantum computation at room temperature~\cite{qbrilliance}. Hence, it is sensible to assume that quantum computers may soon become widely available for individuals and organizations. Analogous to how mobile Artificial Intelligence (AI) accelerators gave rise to performance-critical intelligent applications at the edge relying on Computer Vision (CV) or Natural Language Processing (NLP), the emergence of mobile QPUs will pave the way for a new class of intelligent applications that, for example, benefit from efficiently solving complex optimization tasks. Still, even with centralized approaches, the complexity of developing hybrid applications significantly obstructs advancements in quantum research. Practitioners must consider classical and hybrid components, choose the appropriate hardware, and manually manage the orchestration.
\begin{table}[h]
\centering
\caption{Summary of the Status Quo and our Vision}
\label{tab:vision}
\resizebox{\columnwidth}{!}{%
\begin{tabular}{llll}
\toprule
Category          & Current Situation                                                                                 & Vision                           &  \\ \toprule
Accessability     & \begin{tabular}[c]{@{}l@{}}Restrictive, centralized,\\ consolidates resources\end{tabular} & Optionally centralized or decentralized &  \\ \midrule
Complexity &
  \begin{tabular}[c]{@{}l@{}}Requires selection of components (e.g.,quantum \\ computers), multiple vendors and frameworks\end{tabular} &
  \begin{tabular}[c]{@{}l@{}}Heterogeneous components abstracted \\ behind homogenous interfaces\end{tabular} &
   \\ \midrule
Applications      & Restricted to cloud, high network latency                                                  & Unrestriced, devices co-operation       &  \\ \midrule
Service Guarantees &
  \begin{tabular}[c]{@{}l@{}}Requires assessment end-to-end \\ performance of manually stitched applications\end{tabular} &
  \begin{tabular}[c]{@{}l@{}}SLOs, for single hybrid application.\\ Automates orchestration resource allocation\end{tabular} &
   \\ \midrule
Workflow &
  \begin{tabular}[c]{@{}l@{}}Must integrate individual classical and\\ quantum service providers manually\end{tabular} &
  \begin{tabular}[c]{@{}l@{}}Freely composable workflow of \\ quantum and classical components\end{tabular} &
   \\ \midrule
Resource Locality & Cloud only                                                                                 & Edge-Cloud Continuum                    &  \\ \bottomrule
\end{tabular}%
}
\end{table}
To this end, this work aims to encourage quantum computation research at the edge. We describe key challenges and possible solution approaches for distributed hybrid classical-quantum architectures. Specifically, the focus is on task partitioning to illustrate the potential of drawing from an edge-cloud continuum and to explain the intricate interplay of numerous classical and quantum components the system designs stitch together into one cohesive unit. Ultimately, our vision is for a distributed hybrid platform that can automate the end-to-end process of orchestrating hybrid applications to emerge. Then, as conceptualized by \Cref{fig:vision} and summarized by \Cref{tab:vision}, practitioners can dedicate their time solely to implementing the core logic of their applications and algorithms. 

To concretize and increase the intuition of the abstract concepts of our work, we use a running example of implementing a distributed hybrid platform for Mobile Augmented Reality (MAR). Our choice is sensible, considering how conceiving methods to accommodate the demanding and complex requirements of MAR applications with edge computing generalize well to arbitrary applications and is an active area of research~\cite{siriwardhana2021survey}. Moreover, a platform hosting MAR applications must reliably serve numerous intelligent tasks that can benefit from QPUs. For example, a navigation system for cyclists to safely navigate urban areas by preventing them from crashing into approaching vehicles requires object detection and tracking models from a distributed camera network~\cite{rausch2021towards}.

We summarize our contributions as
\begin{itemize}
\item Highlighting the opportunities of incorporating QPUs into the edge-cloud continuum and how the distinct properties of QPUs introduce novel challenges.
\item Introduce the architecture for a distributed hybrid platform with a variable composition of heterogeneous quantum and classical nodes. 
\item Demonstrating the feasibility of extending partitioning methods for classical Deep Neural Networks (DNN) with quantum embeddings by implementing, evaluating, and open-sourcing a proof of concept (PoC) of a splittable hybrid classical-quantum neural network\footnote{https://github.com/rezafuru/QuantenSplit}.
\end{itemize}

\Cref{sec:background} establishes the preliminary background and related work. \Cref{sec:qcrole} argues how future work on Quantum Edge Computing (QEC) can benefit from existing work on Classical Edge Computing (CEC) by drawing parallels to AI accelerators and their role in Edge Intelligence. The rest of this work is structured to progressively introduce the components of our envisioned platform from a top-down perspective. \Cref{sec:absmodel} presents the core high-level serverless abstraction model for client programmers. \Cref{sec:warmstart} extends the model to provide first-class support for warm starting. \Cref{sec:disarch} focuses on lower-level system challenges for a distributed hybrid inference engine. \Cref{sec:quantensplit} describes and evaluates our PoC for a hybrid classical-quantum D(Q)NN for split inference. \Cref{sec:conclusion} concludes our work.
\section{Background \& Related work} \label{sec:background}
Cloud-centric platforms have paved the way to make cost-efficient and large-scale applications accessible to the public.
However, the emerging edge-cloud continuum accentuates the drawbacks of centralized architectures.
Promising application paradigms, such as Edge Intelligence \cite{deng2020edge}, heavily rely on the edge-cloud continuum and require autonomous management over the large and heterogeneous system.
We argue that quantum computers can improve the quality of existing applications and pave the way for novel ones.
The success of these applications is tied to available platforms that need to support developers in designing, writing, testing, deploying, and managing them.
This section introduces concepts fundamental to our architectural vision and summarizes related work.
\subsection{Orchestration} \label{subsec:orchestration}
The services of centralized platforms that provide access to quantum computers can be combined with classical applications (i.e., Amazon offers event-based processing for their quantum service).
Hence, practitioners and researchers must build hybrid applications by combining separate quantum and classical components.
Worse, they are burdened with selecting different QPU technologies, devices, and compilers~\cite{garcia2021quantum, quetschlich2022predicting}. 
\subsubsection{Orchestration of Quantum Applications}
Although quantum applications consist of classical and quantum components, they follow the same framework as classical computing in dividing orchestration into two distinct procedures~\cite{Weder2021_OrchestrationsInSuperposition}.
First, workflow technologies manage control flows.
Second, provisioning technologies handle the deployment of application components. 
Hence, we can reduce infrastructure complexity by extending existing systems to support quantum applications. Wild et al. present Tosca4Q, which extends Tosca to support workloads relying on quantum computers~\cite{tosca4qc}.
Weder et al. introduce Quantum Application Archives (QAAs), allowing orchestration methods to treat quantum applications as self-containing entities~\cite{Weder2021_OrchestrationsInSuperposition}.
Later, Leymann et al. propose extending QAAs through a marketplace, with an architecture for a collaborative software platform to consider the development process~\cite{leymann2020quantum}.
\subsubsection{Quantum Platforms} 
Several cloud offerings provide access to quantum computing as a service. However, it is still challenging to integrate managed quantum services cohesively into classical applications.

Garcia-Alonso et al. \cite{garcia2021quantum} present their proof-of-concept implementation of a Quantum API Gateway recommending a quantum computer target to run a given quantum application for Amazon Braket. Additionally, Beisel et al.~\cite{beisel2023quokka} propose \textit{Quokka}, a microservice-based framework to model and deploy quantum workflows.
The authors propose a set of microservices that model the typical quantum workflow based on \textit{Variational Quantum Algorithms (VQAs)}~\cite{vqas}.
This workflow comprises circuit generation, execution, error mitigation, objective evaluation, and parameter optimization.
The approach fully decouples each part of pre-processing, executing, and post-processing, allowing a flexible workflow definition.
Further, Salm et al. \cite{salm2020nisq} present a concept that automatically handles the analysis of quantum algorithms and the selection of quantum computers.
Grossi et al. \cite{grossi2021serverless} build a prototypical platform inspired by Serverless Computing through which quantum developers can deploy their applications.
They employ a scheduler that focuses on queue management and result retrieval.
Leymann et al. \cite{leymann2020quantum} propose an architecture for a collaborative software platform for quantum applications that encompasses the development process and deployment aspect through a marketplace for quantum applications.

The systems so far have shown how platforms can enhance collaboration, improve the development of applications, and simplify deployment aspects (e.g., dynamically selecting a quantum computer).
However, they only considered cloud systems and focused on software development.
We see that complementary to our work as advances in these fields can enrich development aspects of our envisioned hybrid application platform.
For example, an adoption of Tosca4Q \cite{tosca4qc} to model serverless applications \cite{wurster2018modeling} can make our envisioned serverless-based platform more accessible.
Still, it requires a platform to manage hybrid applications across the edge-cloud continuum to make quantum applications accessible. 
\subsection{Serverless Edge Computing} \label{subsec:serverlessec}
A key problem of edge-cloud applications, such as edge intelligence, is the autonomous orchestration of edge-cloud applications \cite{deng2020edge}.
Manual management is infeasible in these large-scale and geo-distributed infrastructures, so a platform that can autonomously manage application deployments is required.

Serverless Edge Computing is the natural extension of Serverless Computing that abstracts the underlying infrastructure and transparently deploys applications packaged as functions across the system \cite{aslanpour2021serverless, nastic2022serverless}.
We argue autonomous management and simplified application development and deployment can be enablers of the emerging quantum computing paradigm.
Nguyen et al. \cite{nguyen2022qfaas} present a holistic serverless platform that supports classic, quantum, and hybrid applications. Conversely, we envision a platform that spans the edge-cloud continuum and manages application deployments across heterogeneous infrastructures. 
The increased complexity stems from the composed applications and the sophisticated and fine-grained monitoring, as \Cref{sec:absmodel} will discuss.
\subsection{Task Partitioning} \label{subsec:partitioning}
Task partitioning in classical edge computing and quantum computing are distinct research areas that address orthogonal problems. Nevertheless, they share a common idea in dividing a task into subtasks executable by isolated compute nodes. 

Although we can draw from existing quantum computing partitioning methods to aid schedulers in their placement strategies, we must consider the unique properties of QPUs before we can generalize them to classical methods.

Partitioning in classical edge computing concerns distributing load for resource efficiency~\cite{partitioningclassical}. In quantum computing, splitting tasks between classical and quantum nodes is necessary for near-term applications to cope with the limitations of NISQ devices~\cite{nisqbitter}, and most common hybrid classical-quantum splitting patterns assign fixed roles to components~\cite{partitioningquantum}. Additionally, patterns for quantum computation are typically conceived to execute a particular class of algorithms and do not consider applications where a quantum algorithm is just one of several subtasks. 

In contrast, as the limitations of quantum computation  will gradually diminish, a platform should be able to accommodate new emerging patterns. For example, IBM's 433-qubit QPU announced in 2022 will soon be superseded by an 1'121-qubit system~\cite{ibmroadmap}. 
Further, for near- and long-term QPUs, the platform should dynamically adjust the workload between quantum and classical nodes according to target SLOs, internal (e.g., load), and external conditions (e.g., bandwidth). 
\subsubsection{Variational Quantum Algorithms} 
\textit{Variational Quantum Algorithm (VQA)} is a generic framework for optimizing the parameters of a quantum circuit on a classical computer~\cite{vqas}. Depending on the target task, we can derive more specific algorithms, such as \textit{Variational Quantum Eigensolver} for approximating the lowest eigenvalue of a matrix~\cite{vqes} or \textit{Quantum Approximate Optimization Algorithms (QAOAs)} to approximate the solution of a combinatorial optimization problem~\cite{qaoas}. Another notable instance of VQAs is \textit{Quantum Neural Networks (QNNs)} which aim to improve the representation of classical neural networks with embeddings in the Hilbert space. Note, in literature, the distinction between VQAs, Quantum Machine Learning (QML), and QNNs is blurry; thus, for clarity, we refer to QNNs as models that are built and trained for typical ML tasks (e.g., Feature Extraction, Regression, or Classification).
\subsubsection{Warm-Starting} 
The term \textit{warm-starting} is ambiguous due to its widespread usage in classical and quantum computing. For example, it may refer to techniques that reduce resource usage in machine learning and optimization~\cite{ash2020warm}. Contrastingly, in classical serverless computing, warm-starting typically relates to methods for preparing execution environments, such as reusing running containers~\cite{manner2018cold}. This work considers warm-starting a general strategy for partially computing or preparing a quantum algorithm's output on auxiliary devices. Notably, warm-starting methods are not limited to classical-to-quantum and may be quantum-to-quantum or quantum-to-classical.

Hybrid classical-quantum systems can benefit from various warm-starting methods that utilize previously obtained solutions, approximations, or trained models~\cite{truger2023warm}.
For example, following the assumption that optimal variational parameters for similar problem instances solved with VQAs are in proximity, parameters can be transferred between instances as an initial point to warm-start from and improve upon~\cite{galda2021transferability}.
Moreover, approximations that are cheaply generated by efficient classical algorithms can be utilized to initialize quantum circuits with a quantum state biased towards potential solutions rather than starting from a neutral initial state~\cite{egger2021warm}.
On the other hand, QNNs can benefit from pre-trained models through transfer learning, i.e., adapting a classical or hybrid model trained for a general task and training it further to tackle a similar or more precise subtask~\cite{mari2020transfer}.
Evidently, as these warm-starting methods comprise a source algorithm from which information is drawn and a target algorithm that is enhanced with it, warm-starts may indicate potential ways of distributing both classical and quantum computational efforts in the continuum.
\subsubsection{Depth and Widthwise D(Q)NN Partitioning}
Depthwise refers to splitting a large DNN between vertical layers, i.e., it results in sequentially dependent partitions and does not increase parallelization. In classical computing, depthwise partitioning methods commonly aim to facilitate resource efficiency by allowing multiple devices to process a request across the continuum~\cite{furutanpey2022adaptive}. In quantum computing, depthwise partitioning may refer to stacking QNN circuits to mitigate cascading gate errors and accumulating noise during training~\cite{bondarenko2020quantum}.

Conversely, widthwise partitioning facilitates parallelization by horizontally splitting a layer or circuit. Quantum circuit cutting concerns addresses some limitations of NISQ devices by partitioning larger circuits into several smaller subcircuits~\cite{bechtold2023investigating}. Classical widthwise DNN partitioning is less common since AI accelerators parallelize the execution of layers. Still, in highly constrained environments without access to server-grade hardware, methods such as parallelizing filter computation of convolutional layers across devices are sensible~\cite{zhao2018deepthings}.

We identify classical depthwise and quantum widthwise partitioning as viable methods for an essential component of our envisioned platform and will detail them in \Cref{sec:disarch}.

\section{Quantum Computing and Edge Intelligence} \label{sec:qcrole}
We argue that research on Quantum Edge Computing (QEC) should directly extend Classical Edge Computing (CEC) rather than foster isolated communities. In particular, we find that future work on QEC can learn valuable lessons from existing work on Edge Intelligence (EI).  Intelligent tasks commonly refer to solving problems that a program with classical control structures cannot compute tractably or with sufficient precision. EI leverages advancements in specialized hardware for constrained devices (e.g., AI accelerators) to push the computation of such tasks with modern AI methods (e.g., DNNs). Similarly, QEC can leverage the advancements in (mobile) QPUs to compute numerous tasks tractably (e.g., optimization problems) and with higher precision. 

To identify where we can utilize prior experiences, this section draws parallels between the rise of energy-efficient AI accelerators and the challenges of integrating QPUs. Note that, despite the existence of a broader definition for ``AI”, literature in edge computing typically uses AI as synonymous with classical Machine and Deep Learning. To avoid ambiguity, we will adhere to the same convention. Moreover, for brevity, we refer to AI accelerators covering a broad class of chips suitable for mobile devices that can efficiently execute vendor-specific compiled computational graphs of DNNs as Tensor Processing Units (TPUs)~\cite{li2020survey}.
\subsection{Parallels between Mobile QPUs and TPUs} \label{subsec:aipar}
QPUs can efficiently solve some problems intractable for classical computers (e.g., integer factorization and discrete logarithms~\cite{Shor1994factoring}). Notably, QPUs can utilize entanglement and superposition, two properties unique to quantum computing. Conversely, although adding a TPU is formally incomparable to the speed-ups and precision QPUs can provide, in practice, TPUs provide substantial speed-ups for complex tasks DNNs excel at. 

For brevity, we will refer to DNNs specifically designed for resource-constrained devices (e.g., IoT devices) or DNNs compressed with quantization, pruning, or knowledge distillation methods~\cite{modelcompressionsurvey, gou2021knowledge} as ``Lightweight”.
Analogously, we refer to circuits designed to run on more constrained QPUs than the state-of-the-art as ``shallow”  or ``narrow". The reference point is relative and aligns with the assumption that integrated chips in resource-constrained devices have lower capacity chips than contemporary server-grade hardware. 
Similarly to how TPUs are accelerators for lightweight DNNs~\cite{chen2022mobile} to solve CV and NLP problems, mobile QPUs~(MQPUs) will allow constrained devices running shallow quantum circuits to solve optimization problems. 

Conclusively, viewing QPUs as an accelerator for an orthogonal set of intelligent tasks, i.e., tasks impractical for classical handcrafted programs, is reasonable. Nevertheless, QPUs and TPUs can complement each other for non-orthogonal tasks in the near term. For example, the primary motivation of QNNs interestingly differs from the typical advantages of quantum computation. Since the training and inference of DNNs are highly parallelizable, modern classical hardware already efficiently accommodates them. Instead, QNNs can exploit properties of the Hilbert Space to find better representations~\cite{abbas2021power}. Moreover, it has been shown that utilizing entanglement in training data can reduce the limits on learnability imposed by the (classical) no-free-lunch theorem~\cite{Sharma2022qnfl}. Nevertheless, especially for high-dimensional data (e.g., Images), it is challenging to build and train QNNs since methods must map the high-dimensional representation to a low number of qubits. Therefore, currently, hybrid classical-quantum QNNs show more promising results. Unlike pure QNNs and other VQAs, hybrid QNNs consist of parameterized classical and quantum components, and first apply
classical layers to find a suitable representation before encoding the features to a quantum state and passing them to a quantum circuit~\cite{mari2020transfer, lloyd2020quantum}.\subsection{Quantum for and on the Edge}
In their seminal work, Deng et al. \cite{deng2020edge} proposed distinguishing Edge Intelligence between \textit{AI for Edge} and \textit{AI on Edge}. Analogous to their work, we suggest that future research in QEC should be categorized in Quantum on and for Edge.
\subsubsection{Quantum on Edge (QoE)} addresses the challenges of building systems for distributed applications that rely on QPUs. Research in this category can focus on integrating existing quantum systems into the continuum, such that classical resources across a hierarchical network can aid in reducing wait times and load on QPUs. Alternatively, the focus may be on generalizable approaches for extending existing classical systems to increase the solution quality with QPUs. \Cref{sec:warmstart} and \Cref{sec:disarch} introduce ideational methods for both subcategories to facilitate this direction. 
\subsubsection{Quantum for Edge (QfE)} focuses on solving optimization problems in edge computing. Unlike QoE, it is not fcomplementarity to the classical AI counterpart. Rather, quantum researchers should aim to outperform AI methods. Existing classical methods could be utilized to warm-start quantum optimization. Moreover, they can serve as valuable benchmarks, marking a threshold quantum algorithms must cross before they are viable in practice. 
It is worth noting that quantum algorithms have been shown to outperform classical optimization methods for specific problem instances, which can be identified with a high probability prior to running the algorithm~\cite{Moussa_2020}.

\section{An Architecture for a Distributed Hybrid Compute Platform} \label{sec:absmodel} 
This section introduces the envisioned architecture illustrated in \Cref{fig:basearchitecture}. We first provide a high-level model of an edge-cloud continuum and relate it to our running example of a MAR platform that aims to implement our architecture. Then, we describe the individual components and how clients interact with them so that the remaining sections can reference the architecture.
\begin{figure}[ht]
    \centering
    \includegraphics[width=\columnwidth]{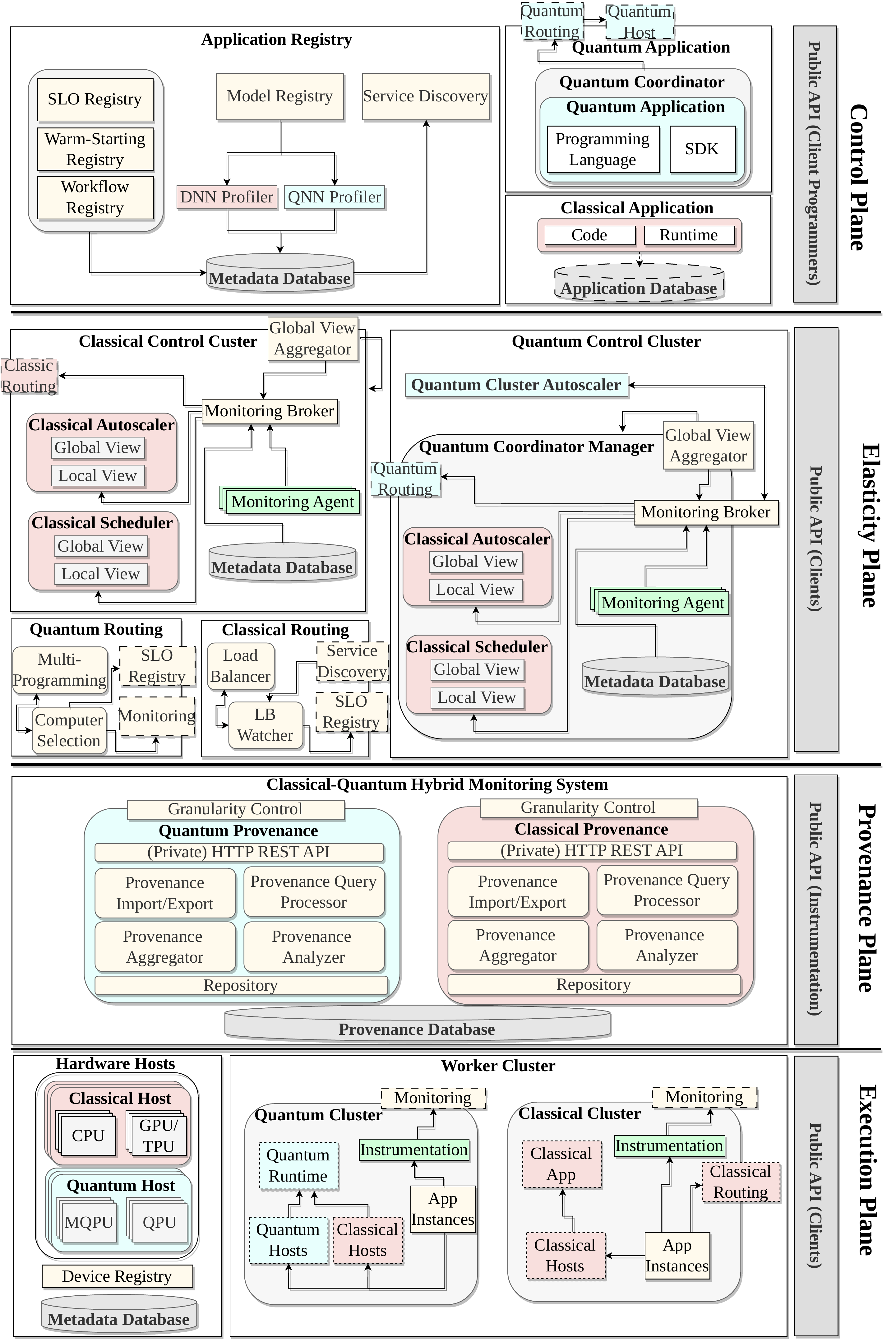}
    \caption{Our envisioned Architecture}
    \label{fig:basearchitecture}
\end{figure}
We do not claim that the components we introduce in our architecture are sufficient for a hybrid platform. Rather, we believe it forms a foundation for future work as a starting point for a new research directive to advance quantum and edge computing.
\subsection{Resource and Scenario Formulation}
Our conceptual MAR platform distinguishes between system designers, client programmers, and clients. The system designers seek methods to build a platform that allows client programmers to deploy applications or experiments. Client programmers write and deploy MAR applications and have varying requirements. Clients are devices that access applications by issuing requests.  Our objective is an architecture that maximizes the rate of advancements in quantum applications and algorithms by \textit{unconditionally} minimizing the complexity for client programmers at the expense of system designers and platform providers.
\subsubsection{Service Level Objectives and Agreements}
With SLOs, clients can specify quantifiable metrics according to their application's requirements that a platform should uphold. Typically providers bundle SLOs together as Service Level Agreements (SLAs). Nevertheless, we do not require a distinction between SLOs and SLAs since the latter is simply a constructible contract of the former. Guaranteeing SLOs is difficult even in purely classical serverless (edge-)cloud systems~\cite{seo2021slo, nastic2020sloc, raith2022mobility}. The following proposes solutions for conceptualizing an architectural platform vision with the additional consideration of integrating QPUs. \Cref{sec:qcrole} stressed the importance of extending existing methods to accommodate QPUs instead of conceiving novel methods unaware of recent advancements in CEC. Therefore, we draw from ongoing research in classical computing from a systems perspective, intending to extend existing state-of-the-art classical methods instead of competing with them. For the remainder of this work, we refer to a component or service as \textit{SLO-aware} if it has access to the SLO registry and implements a corresponding decision mechanism.
\subsubsection{Resource Assumptions on the Edge-Cloud Continuum}
Resources are spread across three domains: edge, fog, and cloud. There is no universally agreed-upon definition for the edge and fog domain. Hence, we remain as general as possible without compromising the practicability of our architecture. Defining the fog domain as clustered computational nodes significantly nearer to client devices than the closest cloud data center is sufficient for our purposes. The edge domain comprises the clients and nearby SLO-aware load balancers~\cite{bogdanov2018fast}. \Cref{tab:resourcesdomain} summarizes each domain's properties and the available resources to which our MAR platform has access. In particular, for QPUs, we refer to the properties of current or expected near-term devices. While once QPUs become mature enough, they may be as accessible as classical hardware, system designers will need to address the near-term limitations of QPUs. 
\begin{table}[h]
\centering
\caption{Summary of Resources Across Domains}
\label{tab:resourcesdomain}
\resizebox{\columnwidth}{!}{%
\begin{tabular}{c|l|c|c|c|c}
Domain                 & \multicolumn{1}{c|}{Dominating Cost}                                                                                 & Type & Grade  & Capacity  & Notable Limitation  \\ \hline
\multirow{2}{*}{Edge}  & \multirow{2}{*}{\begin{tabular}[c]{@{}l@{}}Energy - Client \\ may not warm start\end{tabular}}                       & CPU  & Mobile & Low       & Heterogeneity       \\ \cline{3-6} 
                       &                                                                                                                      & TPU  & Mobile & Very Low  & Lightweight DNNs    \\ \hline
\multirow{3}{*}{Fog}   & \multirow{3}{*}{\begin{tabular}[c]{@{}l@{}}Load - Requests may \\ have to be routed to \\ the cloud.\end{tabular}}   & QPU  & Mobile & Low    & Circuit Width/Depth \\ \cline{3-6} 
                       &                                                                                                                      & CPU  & Server & High      & Outages             \\ \cline{3-6} 
                       &                                                                                                                      & GPU  & Server & High      & Outages             \\ \hline
\multirow{3}{*}{Cloud} & \multirow{3}{*}{\begin{tabular}[c]{@{}l@{}}Communication - \\ Remote data centers\\ incur high latency\end{tabular}} & QPU  & Server & Medium    & Availability        \\ \cline{3-6} 
                       &                                                                                                                      & CPU  & Server & Unlimited & Interference        \\ \cline{3-6} 
                       &                                                                                                                      & GPU  & Server & Unlimited & Interference       
\end{tabular}%
}
\end{table}
We distinguish between mobile- and server-grade hardware. The former can only execute a subset of the latter. Capacity refers to the size of a model and how many instances we can spawn on demand. \textit{Unlimited} means we can host as many instances of the largest off-the-shelf models as necessary. \textit{High} implies some trade-off between the size and the number of instances of a model where it is possible to exhaust the available capacity when the number of concurrent client requests exceeds a certain threshold.  \textit{Medium} and \textit{Low} suggest additional limitations to the hardware, e.g., how the circuit depth and the number of qubits on NISQ devices limit the size of a QNN. In our running example, the platform providers aim to place QPUs near several major cities. Therefore, they purchased numerous mobile-grade QPUs (e.g., the announced diamond-based QPUs of Quantum Brilliance~\cite{qbrilliance} or SaxonQ~\cite{saxonq}) that they expect to have sufficient capacity for most models. Since they cannot feasibly host server-grade QPUs (e.g., IBM's planned 1’121-qubit system~\cite{ibmroadmap}) in every supported location, the requests that require larger circuits are either forwarded to a remote cloud data center or processed with partitioning methods we will elaborate on in \Cref{sec:disarch}. Lastly, we label the capacity of edge TPUs as ``very low“ since they can typically only host one lightweight DNN in memory. Contrastingly, in the fog and cloud domain, we can host multiple instances of the same DNN on virtualized GPUs.  A notable limitation for a resource may not be exclusive to a single domain but is a concern system designers should be particularly conscious about in that domain. Heterogeneity refers to the numerous chip architectures with varying constraints. Lightweight DNNs and Circuit Width/Depth refers to the limitation of TPUs and MQPUs to execute larger models. Interference refers to the performance degradation from serving virtualized resources to multiple clients~\cite{nathuji2010q}.
\subsection{Architecture Planes and Components}
The architecture differentiates between four planes according to their function and intended interaction with other components, clients, or client programmers. Each plane exposes private or public APIs. Private APIs are only accessible by internal components, whereas public APIs are accessible by external entities, i.e., clients and client programmers.
\subsubsection{Execution Plane}
Application instances run on the \textit{execution plane}. It consists of \textit{Public APIs}, \textit{Hardware Hosts}, and \textit{Worker Clusters}.  The public APIs allow clients to interact with the application via access points. 

The \text{hardware hosts} are subdivided according to the supported application types, i.e., into classical and quantum nodes. The quantum hosts are further subdivided according to their \textit{grade}, i.e., QPUs are server-grade and MQPUs mobile-grade. The classical hosts additionally subdivide to consider AI workload. Each node is registered by the \textit{device registry} that associates data describing their capacity (e.g., memory size, VRAM, qubits) and stores it into the \textit{metadata} storage. The metadata storage is a highly-available key-value storage (e.g., ETCD used in Kubernetes) accessible by other components. Especially for the continuum, where nodes can arbitrarily join and leave the system, the metadata must be highly available and not remain stale to measure the system’s overall capacity accurately. 

The \textit{Worker Clusters} consists of at least one hardware host and represents the application environments that may rely on one or multiple hardware nodes. Since quantum and classical environments are separate, it is necessary to distinguish between classical and quantum worker clusters. Naturally, quantum applications will require classical workers for auxiliary tasks (e.g., pre-processing, measurement). Once a worker completes a task, a classical worker is responsible for forwarding the result. The result is forwarded to another worker cluster for distributed applications that partition the task on multiple devices. The result is forwarded to the client for monolithic applications or the last partition of a distributed application. Instrumentation tools send telemetry data to the monitoring system by accessing private APIs of the Provenance Plane. 
\subsubsection{Provenance Plane}
The \textit{Provenance Plane} encapsulates a highly available distributed Provenance database through which real-time monitoring data is stored and shared. It consists of a \textit{Quantum Provenance} and a \textit{Classical Provenance} system, offering private APIs to access and store telemetry data. It is crucial to conceive methods that consider the intrinsic properties of QPUs, to create reliable and predictable systems for quantum workloads. Notably, error rates vary depending on a QPU’s current state, i.e., exclusively collecting classically relevant data (e.g., load) for quantum and hybrid applications is insufficient for SLOs with solution quality targets. Classical monitoring for a platform deploys instrumentation tools (Execution Plane)  alongside the application to collect telemetry data on workload trends and resource usage of function instances (e.g., load, CPUs, GPUs, I/O). Orthogonally, a quantum provenance system gathers information on the state of QPUs to analyze errors, such as properties of Quantum Circuits, QPUs, Compilation, and Execution, as proposed by Weder et al. ~\cite{weder2021qprov}. A platform can utilize provenance data for error mitigation and to aid schedulers with upholding SLOs by assessing the currently expected solution quality. Platforms that support hybrid applications should integrate quantum provenance with classical instrumentation to form one cohesive monitoring system for simplified access to various heterogeneous devices. 

The objective of a monitoring system is to collect the minimal data necessary for informing schedulers to uphold SLOs. Conceiving hybrid systems is non-trivial, as finding a balance is already challenging for classical edge-cloud and hybrid systems monitoring~\cite{georgiou2018streamsight, raith2022mobility}. A system that trivializes monitoring by aggressively collecting data may ease the task of a scheduler but can cause resource congestion across the edge-cloud continuum~\cite{georgiou2018streamsight}. Conversely, collecting insufficient telemetry data will significantly reduce the system's scalability as it cannot appropriately route requests or scale resources up and down. We argue that a \textit{granularity mechanism} capable of adjusting the frequency of quantum and classical monitoring data according to the current workload’s characteristics is one of the principal challenges that future work must address before a hybrid platform can emerge. Nevertheless, monitoring itself is simply a precondition. The following describes how our architecture supports resource efficiency and elasticity based on available monitoring data. 
\subsubsection{Elasticity Plane}
The \textit{Elasticity plane} is the central organ of the system that decides how to allocate resources and route requests according to client SLOs, metadata, and monitoring data. 
The \textit{Control clusters} are subdivided between Quantum and Classical Control Clusters. The former manages quantum application instances (e.g., scale-out quantum coordinators), while the latter manages classical application instances (e.g., horizontal scaling of applications).
Each control cluster manages a set of worker clusters (see Execution Plane) that are dynamically scaled up or down according to the application instances the cluster can control. Monitoring Agents are responsible for the worker clusters and relay data to the control cluster’s monitoring broker. A \textit{Monitoring Broker} disseminates the data across the components. A \textit{Classical Autoscaler} and \textit{Classical Scheduler} maintain a local and global view of the control cluster’s state.  Classical autoscaler and schedulers exist in Quantum Control Clusters, since Quantum Coordinator Applications (See Control Plane) are classic. The \textit{Quantum Coordinator Manager} supervises the Classical Autoscaler and Scheduler to scale and place the Quantum Coordinator Application (see Control Plane) instances. 

The \textit{Local View} contains fine-grained information about the Worker Clusters (e.g., CPU usage per second). In contrast, the \textit{Global View} contains coarse-grained information (e.g., average CPU usage over an hour) about neighboring control clusters. The coarse-grained data consists of fine-grained local data collected by a Global View Aggregator that periodically publishes a summary, i.e., the global view consists of exchanged summaries of local views. 

The control cluster’s messaging topology and broker partially address the granularity control of monitoring data and permit an elastic control mechanism that can scale the entire system by adequately allocating its limited resources. The implementation of the autoscaler and scheduler is interchangeable, and system designers may experiment with various methods. The \textit{Quantum Cluster Autoscaler} is inspired by the work of Tamiru et al. and Gandhi et al. and is an SLO-aware cluster autoscaler capable of adding and removing Quantum Hosts from worker clusters to process the incoming workload.  

\textit{Classical Routing} is inspired by the work by Raith et al. ~\cite{raith2022mobility} and consists of a \textit{Load Balancer} and a \textit{Load Balancer (LB) Watcher}. The Load Balancer is a high-throughput and low-overhead component that re-directs incoming requests to application instances or other clusters (e.g., because no application instance is running). The Load Balancer Watcher is SLO-aware and periodically refreshes the load balancer’s state to update the decision mechanism. 

The \textit{Quantum Routing} component differentiates itself from Classical Routing, as it considers additional challenges to improve the resource efficiency of quantum hosts. \textit{Quantum Computing Selection} is particularly valuable for the edge-cloud continuum as it introduces further heterogeneity. The selection method is another freely interchangeable component. For example, system designers may opt to use the method proposed by Quetschlich et al.~\cite{quetschlich2022predicting} and replace it once they find a more suitable alternative. \textit{Multi-programming} in quantum computing has a comparable role to \textit{virtualization} in classical computing, i.e., it allows sharing of the resources of quantum computers among multiple circuits~\cite{das2019case}. However, unlike in classical computing, where we can readily select existing mature virtualization methods, multi-programming is more involved and should be considered together during encoding~\cite{ohkura2022simultaneous} and influences quantum computer selection. 
\subsubsection{Control Plane}
The Control plane exposes an API to client programmers to deploy and manage their applications. The public API should allow client programmers to define hybrid applications as workflows without separately deploying classical and quantum parts. The platform analyzes the workflow during the registry and partitions it into classical and quantum applications. The \textit{Application Registry} encapsulates several registries responsible for storing application services, SLO targets, and parameterized models. \textit{The Model Registry} is similar to a model zoo (e.g., torch image models~\cite{rw2019timm}). However, here the models may either be classical neural networks or quantum circuits. In addition, profilers associate static metadata to each model (e.g., number of parameters, circuit depth, layer types). Profiler metadata supplements the schedulers and autoscalers with valuable information to predict resource usage more accurately (e.g., a Swin-Tiny will incur higher usage than a Swin-Base~\cite{swin}). Our architecture supports warm-starting as a first-class citizen, which \Cref{sec:warmstart} will elaborate on. Service Discovery enables Quantum and Classical Routing to locate running application instances.

\textit{Classical Applications} consist of the application code and a runtime that executes the code upon creating an application instance. The architecture does not require any assumptions about the runtime, i.e., the runtime can range from containers to WebAssembly modules. Application may optionally provision arbitrary \textit{Application Databases} (e.g., relational, object-based).

The \textit{Quantum Application} is more complex as it requires the interplay between quantum and classical resources. The \textit{Quantum Coordinator} contains the Quantum Application, implemented by the client programmers. The Quantum Application is written in a particular programming language and an SDK for quantum computing. The Quantum Coordinator coordinates the execution of a quantum application (i.e., quantum circuit). Quantum Applications requires routing capabilities to connect the classical part of the quantum application, that pre- and post-processes input and output, with the quantum part, the QPU that executes the quantum circuit. Moreover, Quantum Coordinator requires Quantum Routing capabilities to select a suitable Quantum Host or to perform Multi-Programming to increase QPU utilization efficiency.

\section{Warm-Starting at the Edge} \label{sec:warmstart}
Warm-starting aligns with the objectives of a distributed hybrid platform, i.e., it facilitates drawing from resources across the continuum. Ideally, applications can pre-process input before passing it to a remote server. Warm-starting methods in the context of quantum computing are categorizable into Classical-To-Quantum (C2Q), Quantum-To-Quantum (Q2Q), or Quantum-To-Classical (Q2C)~\cite{truger2023warm
}. Each category has an input and an output format. For example, C2Q expects classical input, and the output format should suit a quantum algorithm. \looseness=-1

Nevertheless, we argue that system designers must subdivide the categories further to include neural input and output formats, such as Neural-To-Quantum (N2Q) or Classic-To-Neural (C2N), for two reasons. First, neural methods rely on AI accelerators that may not be present or have alleviated energy consumption, i.e., it is indispensable for a scheduler to know hardware properties to hit SLO targets. Second, although the output of a classical DNN is classical, the network weights may be tuned to extracted features tailored for a particular class of algorithms.
\subsection{Current and Future Role of Warm-Starting} \label{subsec:roleswarmstarting}
Currently, a common motivation for warm-starting is to reduce the dependency on QPU time due to cost and limited availability~\cite{truger2023warm}. However, we stress that the importance of warm starting lies in improving the solution quality by combining classical and quantum algorithms. Moreover, once QPUs mature to a point where we can entirely forgo classical computation, the research focus can shift to Q2Q warm-starting, where smaller client devices can partially onload quantum algorithms for resource efficiency. 
For example, warm-starting can be a means to embed performance guarantees of classical algorithms into quantum algorithms and can reduce the amount of training data required for (Q)ML. 

A downside of warm-starting is that it increases the applications' complexity as it introduces more parts that must be managed, i.e., it is crucial to introduce a convenient interface that supports warm-starting as a first-class citizen further to shift complexity from client programmers to platform providers. Notably, warm-starting is chainable. Hence, we can naturally integrate warm-starting methods in the edge-cloud continuum if a platform exposes an interface that resembles the hierarchical properties regarding device capacities in the network.\looseness=-1

The following describes the requirements and proposed solution approaches for an aware hierarchical warm-starting programming model. Hierarchical refers to how the warm-starting methods are composable in a \textit{pipeline} that resembles their resource usage requirements. The proposed interface does not rely on any assumption regarding the availability and limitations of QPUs, i.e., it treats each method in the pipeline as exchangeable building blocks. The current progress of available QPUs is reflected in the client programmers by informing them about the feasibility of their planned warm-starting pipeline. For example, the platform could disable support for Q2Q warm-starting at the network's edge until MQPUs find widespread adoption in end devices, such as smartphones. \looseness=-1
\subsection{Hierarchical Warm-Starting} \label{subsec:slowarmstart}
In our running example, the MAR platform aims to improve resource efficiency with hierarchical warm-starting. The load heavily fluctuates for city-scale applications according to date and time. By chaining warm-starting hierarchically, idle computational resources of edge and fog nodes can be utilized, e.g., to reduce offloading to the cloud.

Warm-starting is a broad term encompassing numerous classes of methods~\cite{truger2023warm}. The challenge is to conceive an interface flexible enough to remain convenient without exposing low-level details, such as manually selecting devices and fallback mechanisms, to the clients. An interface would be maximally flexible if it forces client programmers to define every single step of the execution, where to deploy which part at which node, and to manually configure the quantum executions (e.g., device, compiler) for every method in the pipeline. An interface that does not restrict the configuration space is especially undesirable when considering the heterogeneity of the continuum. Specifically, it would not be sufficient to provide a single configuration for a method, and there is no guarantee for the availability of a particular device configuration at the edge or fog.
Conversely, constraining client programmers exclusively to a list of pre-implemented solutions is counterproductive as it hinders innovation, i.e., they should at least be able to (optionally) provide their warm-starting method. 

To summarize, the responsibility of a platform is to build the infrastructure and provide adequate abstractions to access the resources. A dedicated interface for warm-starting should allow clients to define composable workflows to process a warm-starting pipeline, i.e., client programmers can register pre-processing steps for warm-starts through the control plane from our architecture in \Cref{fig:basearchitecture} that may run on CPUs, TPUs, or QPUs deployed at the client device or fog nodes.  

Consider \Cref{fig:hwarmstarthlvl} for the following example. Three clients execute the same application using the public API of the control plane. The client programmers defined a warm-starting pipeline for their applications. Hence, the pipeline and its methods are placed in the warm-starting registry, and the system decides where to position the models in the continuum based on the profiler metadata. However, the clients request varying target qualities; hence, the platform applies different intensities of warm-starting before sending the task to a quantum cloud vendor. Intensity refers to the expected solution quality of a quantum algorithm with an input processed by a warm-starting method.
\begin{figure}[h]
    \centering
    \includegraphics[width=\columnwidth]{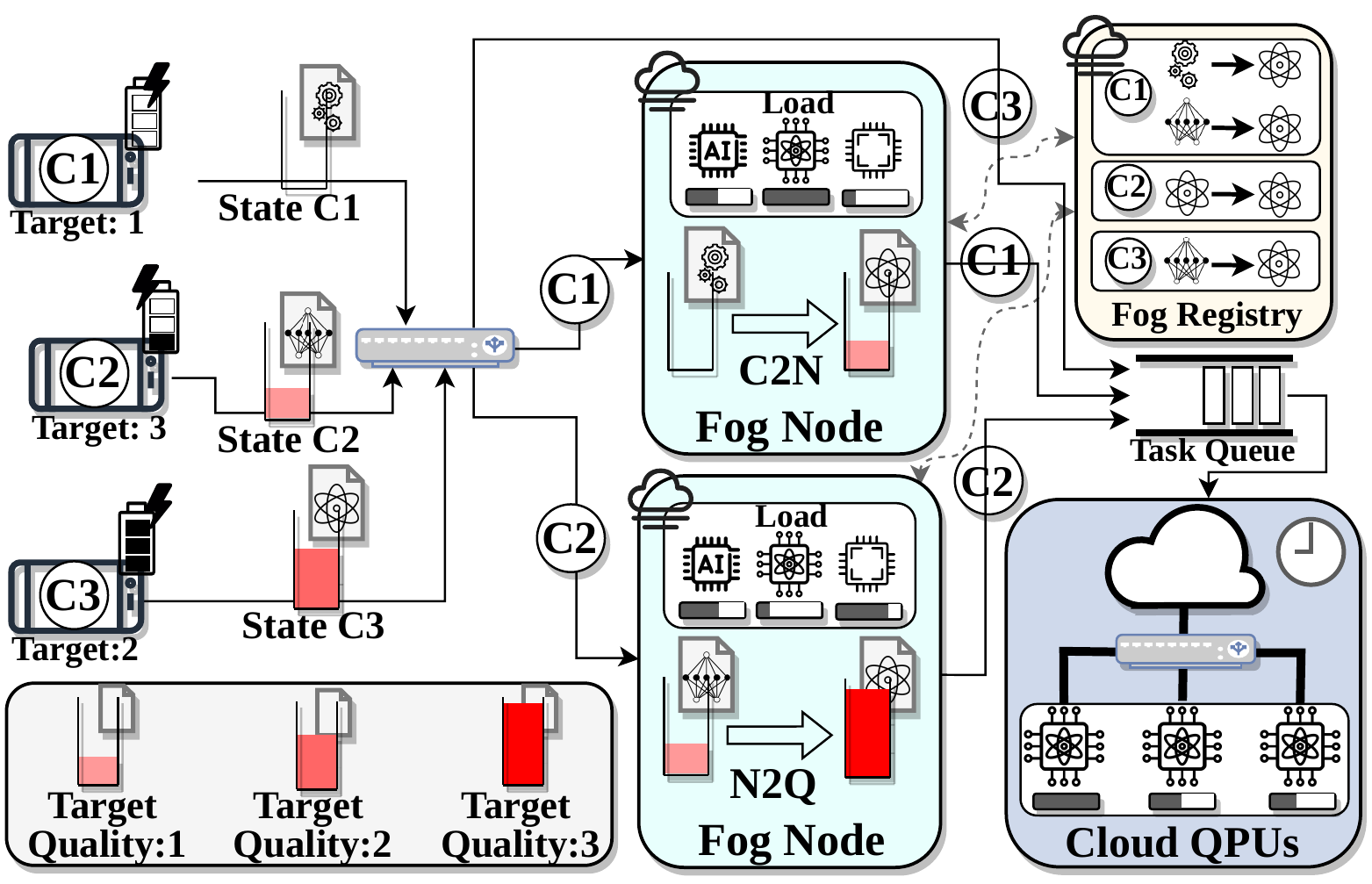}
    \caption{High-level Sequence of Hierarchical Warm-Starting}
    \label{fig:hwarmstarthlvl}
\end{figure}
In \Cref{fig:hwarmstarthlvl},
client C1 cannot achieve any pre-processing in the edge due to energy constraints (indicated by a low battery symbol) and therefore forwards the input in its original classical representation. Since C1 registered at least one pre-processing step at the fog, the load balancer routes the request to a fog node, applying the step that maps to a C2Q warm-start. Conversely, if C1 had the resources to pre-process its input for a target neural network, the fog node would have taken over processing its output further to warm-start the quantum task, resulting in a chain of C2N and N2Q warm-start. Client C2 can perform resource-conscious pre-processing for a C2N warm-start. Still, since it does not achieve the required target quality, the request is routed to a fog node with available QPUs to apply further pre-processing for an N2Q warm-start. Client C3's request is directly routed to the cloud by the load balancer, since C3 had enough onboard resources for pre-processing the task to the target quality without relying on fog nodes.

\section{A Distributed Inference Engine for Hybrid Classical-Quantum Neural Networks}  \label{sec:disarch}
While the last two sections introduced high-level concepts of the architecture, this section focuses on lower-level system designs for platform designers and how client-programmers may implement a distributed hybrid application. We extend our running example and discuss how the MAR platform may support an inference engine with hybrid classical-quantum neural networks.

(D)QNNs are uniquely qualified as a representative application beyond DNN inference to cover how a distributed platform can adapt as the limitations of QPUs are gradually lifted for three reasons. First, a hybrid QNN is composable of parameterized classical and quantum nodes. Contrastingly, other VQAs typically operate non-parameterized classical components exclusively for pre-/post-processing and to optimize the parameters. For a hybrid QNN, there are additional components with no statically predefined role assignments, i.e., we can represent how systems adapt as they progressively replace classical with quantum nodes proportional to availability and advancements in QPUs. Second, we can split a neural network horizontally by layer or cut individual layers vertically and view each partition as an isolated computational graph. Then, a simple inference request (e.g., image classification) can emulate the behavior of a complex task with numerous classical and quantum subtasks that a system must coordinate to compute a single solution. Third, DNN partitioning and collaborative inference are well-established research areas in CEC that consider the heterogeneity of classical hardware and the resource asymmetry between edge, fog, and cloud nodes~\cite{matsubara2022split}. Hence, we can directly extend existing work to determine whether QPUs may improve the quality of classical methods and are not restricted to orthogonal problems infeasible for classical computers.
\subsection{System Challenges} \label{subsec:syschall}
We consider the domain properties summarized in \Cref{tab:resourcesdomain} to describe the challenges system designers may face when implementing our architecture.

Unlike regular business logic, intelligent tasks (e.g., image classification, object detection) rely on specialized hardware, i.e., regardless of whether we include QPUs, the platform must treat inference requests as a workload with distinct characteristics. 
\subsubsection{Volatility}
The scheduler must dynamically adapt to two sources of volatility. First, outages in the fog domain are frequent. Moreover, unlike in the cloud, classical fog resources cannot seamlessly scale horizontally, i.e., requests may have to be routed to the cloud. Second, fog and cloud QPUs may be scarce, and depending on their current state, they may not hit the target solution quality SLO.
\subsubsection{Device Heterogeneity}
While the challenges of heterogeneity of classical components are only tangentially related to the integration of QPUs, minimal consideration regarding the numerous accelerators is necessary. Compilers map classical DNN to computational graphs, and vendors have varying support for operations limiting the available layer types and activation functions~\cite{li2020survey}.
\subsubsection{Task Chaining and Bandwidth Consumption}
To fully draw from the resources on the continuum, we require methods that onload some computation on client-side accelerators. However, mobile devices can typically only host a single network in memory, and swapping out DNN weights from storage incurs significant overhead. Hence, latency-sensitive applications sending subsequent inference requests for different tasks must offload, leaving valuable resources idle. Additionally, when numerous clients compete for limited bandwidth by streaming high-dimensional image data, the limited bandwidth will inevitably lead to erratic response delays.
\subsubsection{Optional Quantum Embeddings}
Although the availability of QPUs is steadily increasing, clients cannot expect the same graceful scaling of classical resources in the cloud currently, i.e., depending on the load, it may not be possible to hit latency SLOs with quantum layers. 
Therefore, the inference engine should be flexible enough to skip computing quantum embeddings for near- and intermediate-term devices.
\subsubsection{Utilizing Mobile Quantum Devices}
Diamond quantum accelerators are expected to be mature enough soon for commercial use~\cite{saxonq, qbrilliance}. However, regardless of how MQPUs improve, analogous to classical hardware, we assume that server-grade hardware will consistently outperform their mobile counterparts. The challenge is to conceive methods that can leverage the advantages of MQPUs, ideally without sacrificing solution quality. 
\subsection{Solution Approach} \label{subsec:solapproach}
The simple solution for accommodating multiple tasks and configurable solution qualities considers a separate DNN for each variation. This is difficult to maintain and inflexible, forcing client programmers to implement and re-deploy entire architectures for each task. Worse, it does not address challenge 3), i.e., applications relying on multiple tasks cannot draw from client resources without a significant latency penalty. Instead, we describe how partitioning methods lead to composable architectures that naturally define small deployable applications.
\subsubsection{Classical Split Computing}
Depthwise split computing addresses challenges 2)-4) and groups layers into partitions. Each partition is a feature extractor, and sequentially applying them is a particular form of hierarchical warm-starting we introduced in \Cref{sec:warmstart}.  Platforms can include pre-trained DNNs in the warm-starting registry of the Control Plane from. Additionally, client programmers may register modules according to their requirements. 
For example, edge devices can optionally perform preliminary feature compression, and fog nodes can apply a small- or medium-sized feature extractor according to solution quality and latency targets. 

To address challenge 2), we require an encoder suitable for constrained end devices (e.g., Smartphones) composed of operations widely supported by the various vendors of AI accelerators. To address challenge 3), the encoder should perform initial feature extraction and find a minimal representation for a sufficient statistic on several downstream tasks to reduce bandwidth consumption. Then, the server can select an interchangeable DNN for additional feature extraction according to the configured latency and accuracy SLO. Lastly, a QNN layer should embed the features to potentially improve the representation before passing it to the classifier. However, to address challenge 4), i.e., to cope with the limited availability of QPUs, the QNN should be optional.
\subsubsection{Quantum Circuit Cutting} 
Quantum circuit cutting addresses challenge 5). The idea is to cut large circuits that require many qubits widthwise into smaller subcircuits requiring fewer qubits~\cite{bechtold2023investigating} by strategically cutting circuit wires~\cite{Peng2019,Brenner2023} and gates~\cite{Mitarai2021,bechtold2023investigating}. 
\Cref{fig:gatevswirecut} illustrates an example in which one wire and two gates are cut.
Wire cutting separates circuit wires through multiple measurements with different observables $O_i$ and subsequent initializations of the qubit to state $|\psi_i\rangle$, while gate cutting substitutes two-qubit gates with varying combinations of single-qubit gates.
The stacked subcircuits in \Cref{fig:gatevswirecut} indicate the generation of multiple variations for each subcircuit.

The widthwise partitioning enables each subcircuit instance to be executed individually on smaller quantum devices, which may be more readily available. Following the execution of these subcircuits, a classical post-processing procedure is applied to recombine the results obtained from the individual subcircuits, ultimately reconstructing the output of the original circuit as a linear combination of the subcircuit results.

This approach facilitates the distribution of quantum circuit computations across multiple QPUs without necessitating quantum communication. As a result, quantum circuit cutting offers the opportunity to harness the power of several smaller MQPUs at the edge, enabling the computation of larger quantum circuits. 
Moreover, it promotes the more flexible placement of quantum circuits across resources of the compute continuum. Additionally, once MQPUs are wildly available, we can leverage them to parallelize the execution of subcircuits. This parallelization can alleviate the overhead associated with each cut, significantly improving computational efficiency and scalability. \begin{figure}[ht]
    \centering
    \includegraphics[width=\columnwidth]{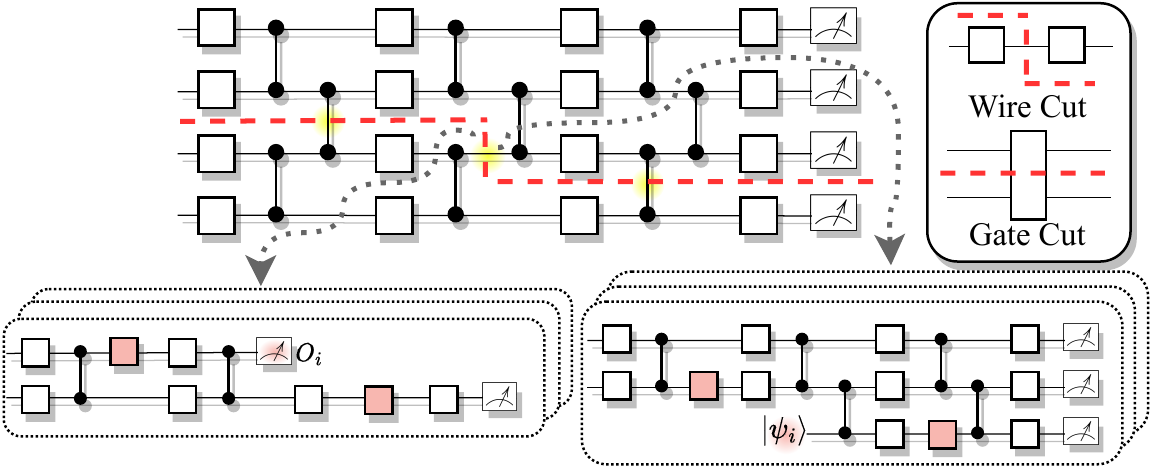}
    \caption{Circuit Cutting Basics}
    \label{fig:gatevswirecut}
\end{figure}
\subsubsection{Inference Flow} \label{subsubsec:infflow}
To provide intuition on how our proposed approaches can serve client requests with varying requirements, consider the flow of inference requests in \Cref{fig:infflow}. For simplicity, we restrict the example to image classification. However, the image classification tasks may be different. For example, one client classifies artwork in a museum to retrieve a description using a virtual tour guide. Simultaneously, another client is interested in retrieving descriptions of the local fauna.
\begin{figure}[ht]
    \centering
    \includegraphics[width=\columnwidth]{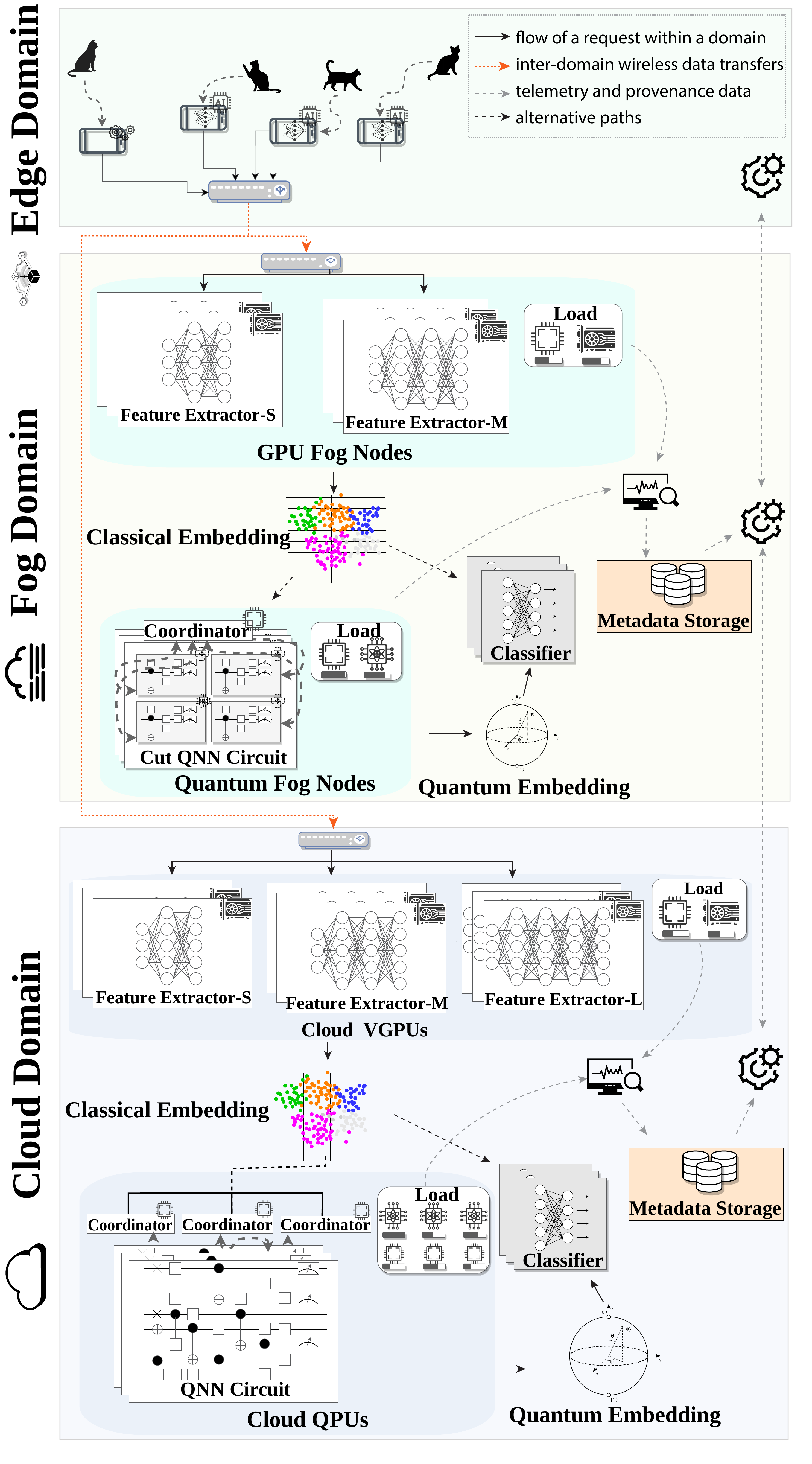}
    \caption{Inference Engine Request Flow}
    \label{fig:infflow}
\end{figure}
Clients with AI accelerators apply neural compression methods for preliminary feature extraction. Solid black lines represent the flow of a request within a domain. The dashed red lines represent inter-domain data transfers. The dashed gray lines represent the telemetry and provenance data collected adjacent to inference requests by a runtime spanning all domains.
The runtime is detailed by the Elasticity Plane of our architecture in \Cref{fig:basearchitecture} and collects telemetry data to periodically update the SLO-aware load balancer to perform informed decisions based on the state of each participating node and client configurations. Dashed black lines represent an alternative path,  i.e., one request flows only to one of the choices. 

An edge load balancer routes the request to the load balancer of a fog cluster. Depending on load and client requirements, the request is routed to a Fog GPU node or the cloud. After feature extraction, a load balancer decides whether the classical embedding should be passed to a QNN before classification. The QNN may be executed on a Quantum Fog Node or sent to a remote cloud provider (e.g., due to privacy or availability). However, based on our assumption, the MQPUs are more constrained than the server-grade QPUs from cloud providers. Hence, circuit cutting methods can aid MQPUs in achieving a target solution quality. A request ends after the classification label is sent as a response to the client.

\Cref{sec:quantensplit} will detail the DNN architecture and demonstrate the viability of our proposed approach. Addressing challenges 1), 5), and introducing runtime components (e.g., SLO-aware schedulers, deployment mechanisms) are other promising research directives but out of scope for this work. Still, the following describes how a platform may realistically prepare and deploy the neural network components.
\subsubsection{Application Preparation and Deployment}
The individual operations of a DNN form a computational graph. Moreover, partitioning methods naturally demarcate a monolithic DNN into connectable vertices. The vertices represent coarse-grained classical layers or QNN circuits, and one or multiple consecutive vertices form one \textit{depthwise partitioned deployment unit} (e.g., a container). Alternatively, we can further partition a vertice to create one \textit{widthwise partitioned deployment unit} (e.g., with circuit cutting). Notably, depthwise methods define isolated compute nodes, which we can transparently combine with widthwise methods, i.e., from an outside view, an adequate abstraction can present a cut circuit as a single coarse-grained layer.

The client programmers may provide hints to the platform via annotations, but the application should be deployable as a single (monolithic) workflow. It is the responsibility of the Control Plane of our architecture to create the deployment units before spawning Quantum and Classical Application instances. From the point of view of the client programmers, they have deployed a single application. However, the runtime system should be aware that the application is split into multiple parts.
\Cref{fig:compgraph} illustrates an example with a computational graph of coarse-grained layers. 
\begin{figure}[h]
    \centering
    \includegraphics[width=\columnwidth]{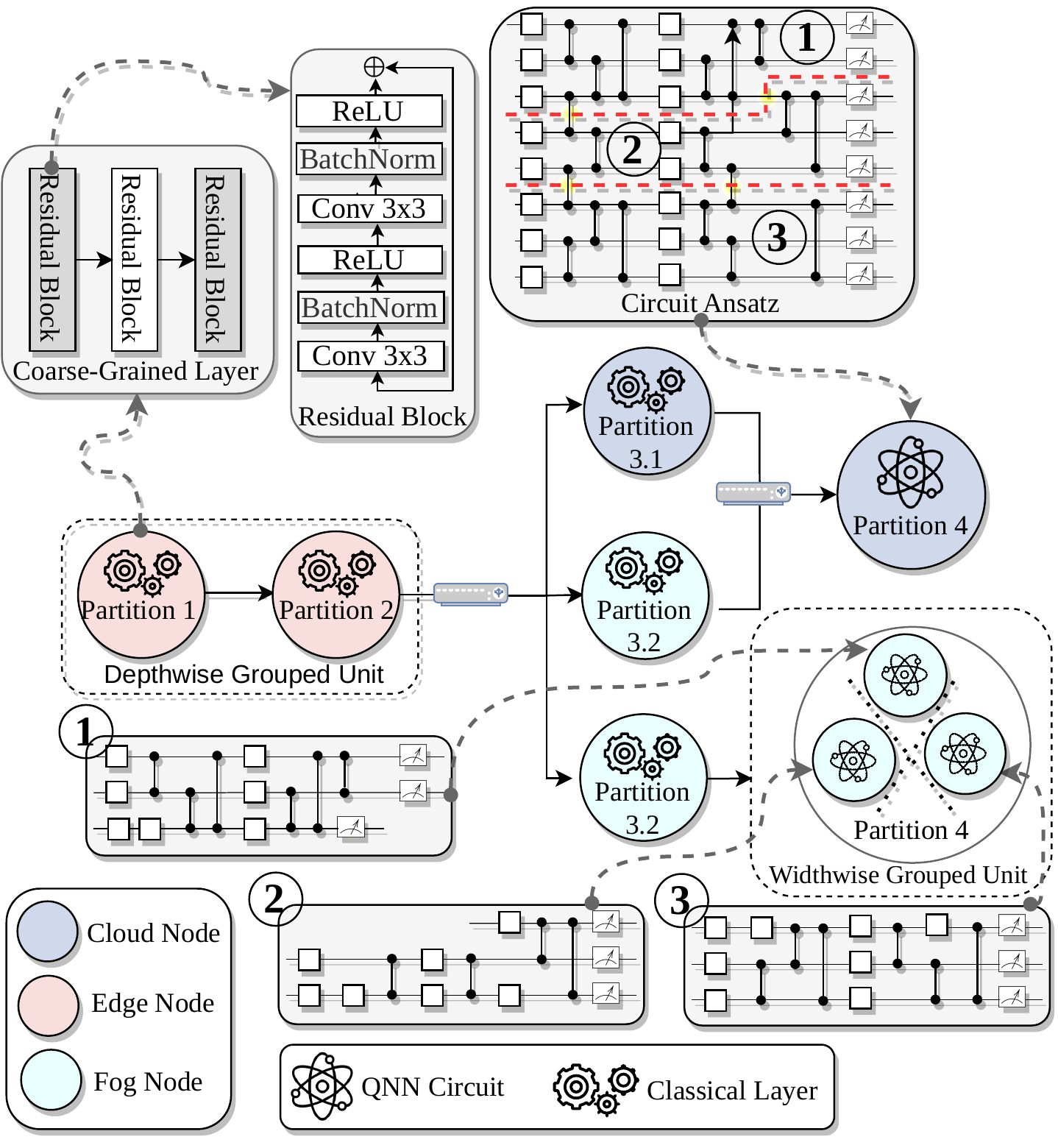}
    \caption{Coarse Grained Deployment Units}
    \label{fig:compgraph}
\end{figure} 
A coarse-grained layer consists (recursively) of finer-grained layers. The nodes are enumerated to indicate the processing sequence, and a subindex indicates a branching path. A node with the same index and subindex implies the same partition deployed on different nodes. Partitions 1-2 are grouped depthwise and will be deployed as one unit on edge devices. Partition 3.2 is deployed on cloud and fog nodes, while Partition 3.1 is a different model deployed exclusively on cloud nodes. For example, Partitions 3.1 and 3.2 could be Feature Extractor L and M from \Cref{fig:infflow}. An SLO-aware load balancer routes the output of Partition 2 to a variation of Partition 3. Lastly, the output of Partition 3 is passed on to one of the instances of Partition 4. Partition 4 is deployed on fog and cloud nodes. However, it is a QNN circuit that a cloud QPU can execute but must be partitioned widthwise (i.e., cut into subcircuits as described in \Cref{subsec:solapproach}) for the constrained MQPUs at the fog nodes. 

\section{Split Inference with Classical-Quantum Hybrid Predictors} \label{sec:quantensplit}
Arguably, demonstrating that we can extend classical DNN partitioning methods to classical-quantum DQNN is a necessary precondition for a distributed hybrid inference engine. Hence, to show the viability of our visions, this section addresses challenges 2)-4) from \Cref{sec:disarch} (and partially 1) with a partitionable neural network architecture where the runtime of the Elasticity Plane can freely decide using a hybrid or a classical predictor.

To encourage follow-up work, we open-source a repository that provides a convenient framework to add configurable experiments with new circuit or model implementations. 
\subsection{Problem Formulation} \label{subsec:qsprobform}
As advocated for in \Cref{sec:qcrole}, we extend a method originally conceived for CEC to QEC instead of disregarding existing work. Specifically, we introduce \textit{QuantenSplit}, a simple modification of the split inference method \textit{FrankenSplit}~\cite{frankensplit} with depthwise DNN partitioning, to support quantum embeddings with QNNs for image classification. 

We choose FrankenSplit since it is not limited to a single head-tail pair, so client devices do not have to swap out head weights whenever two subsequent requests require a different head network. Moreover, FrankenSplit focuses the local resources on bandwidth reduction with a variational feature compression model, i.e., an extension to support QNNs addresses challenges 3) and 4) from \Cref{subsec:syschall}. The method draws from the Information Bottleneck (IB)~\cite{informationbottleneck} principle to achieve considerably higher compression rates than handcrafted codecs without sacrificing predictive strengths.

We omit the formal, conceptual details and instead refer to the original work~\cite{frankensplit}. Here, it is sufficient to consider FrankenSplit as a framework to train a variational autoencoder that is particularly selective about the signals it discards during compression. Notably, attaching different (split) neural networks for multiple downstream tasks to the autoencoder is possible. In other words, it permits a platform to deploy a universal encoder to all participant clients agnostic towards their particular inference requests. 

To determine whether FrankenSplit is generalizable to hybrid classical-quantum QNN predictors, we answer the following question: \textit{Do the highly compressed features of the classical universal encoder generalize to downstream tasks with QNNs?}.  With this, we aim to show the usefulness of our envisioned platform from \Cref{sec:absmodel} and to provide evidence for the viability of our distributed inference engine discussed in \Cref{sec:disarch}.
\subsection{Methodology}
To answer the above question, we must show that we can attach hybrid and classical predictors to a single autoencoder-backbone pair. We consider four datasets resulting in eight predictors, i.e., for each dataset, we attach one hybrid and one classical classification model to the base network.
\subsubsection{Compression and Feature Extraction}
 \Cref{fig:qsarch} illustrates the architecture of the modular neural network. The backbone is a pre-trained split classical neural network that further extracts features. 
\begin{figure}[ht]
    \centering
    \includegraphics[width=\columnwidth]{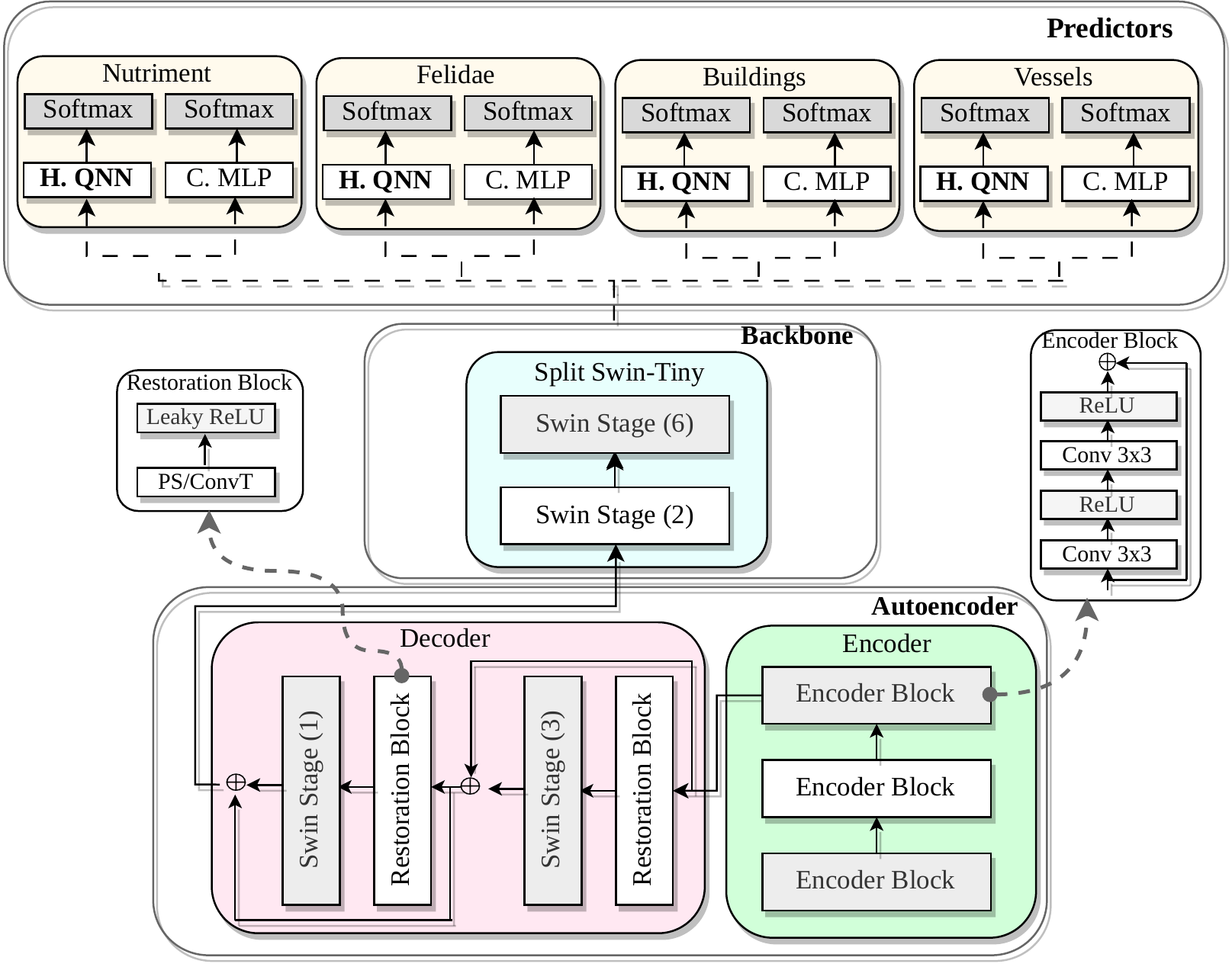}
    \caption{QuantenSplit Network Architecture}
    \label{fig:qsarch}
\end{figure}
It is possible to attach several backbones of varying sizes (or architectures) such that the network is deployable, as depicted in \Cref{fig:infflow}. Nevertheless, we only require attaching multiple (hybrid) classification models to a single pre-trained backbone for our purposes. The encoder only relies on widely-supported operations to deal with device heterogeneity in the edge domain. The number in the bracket denotes the stage depth. The decoder blocks smooths out the quantized features and transforms features to suit the backbone. The split Swin-Tiny backbone is described in \cite{swin}, except the first two stages are discarded. The difference from the original work is that we attach hybrid classification models in addition to the classical ones. 
\subsubsection{Classification Models}
A classical classifier is a simple two-layer Multi-Layer Perceptron (MLP). \Cref{fig:qscircuit} illustrates the Ansatz of the QNN with four qubits and layers.
\begin{figure}[ht]
    \centering
    \includegraphics[width=\columnwidth]{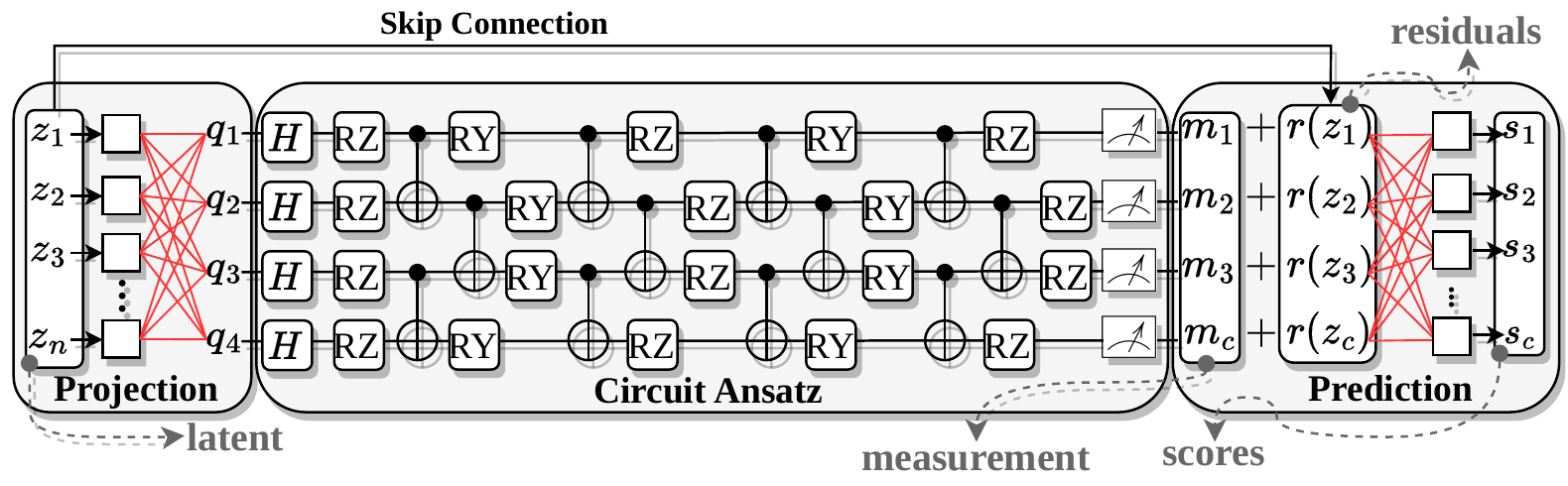}
    \caption{Hybrid QNN with Four Qubits and Layers}
    \label{fig:qscircuit}
\end{figure}
A Hybrid QNN model takes as input the n-dimensional real-valued feature vector  $\mathbb{Z}^n$ and classically projects it to a vector with dimensions equal to the number of qubits. Then, it passes the features as input to the Ansatz. Regardless of circuit depth, it first applies a Hadamard $H$ Gate and a parameterized $Z$-rotation $RZ$ to embed features in the quantum node. Next, it applies a repeating sequence of trainable variational layers. A layer consists of pairwise (shifted) C-NOT gates followed by alternating parameterized $Y$- or $Z$-rotations, i.e., a layer with $Y$-rotation is followed by a layer with $Z$-rotations. The number of layers is a hyperparameter given by the depth of the circuit. Lastly, it passes the measurements to a fully connected layer to output the class scores. The skip connection is inspired by classical Residual Neural Networks~\cite{he2016deep}. Adding the skip connection is configurable as a hyperparameter. 

\subsection{Training and Implementation}
We first separately train the autoencoder on the 1.28 ImageNet~\cite{imagenet} training samples according to the baseline objective function of FrankenSplit. Then, we freeze the parameters and train each attached classification model sequentially using the cross entropy loss function. We use Adam optimization~\cite{kingma2014adam} for the autoencoder and all classification models. Applying widthwise partitioning methods, such as circuit cutting, is out of the scope of this work. A detailed description of the training parameters can be seen in the configuration files of the accompanying repository. 

We did not perform exhaustive hyperparamter tuning or experiments regarding optimizers. We implement our models using PyTorch~\cite{pytorch2018pytorch} and CompressAI~\cite{begaint2020compressai}. The backbones and pre-trained weights are from PyTorch Image Models~\cite{rw2019timm}. For numerical simulations of the quantum circuits, we use PennyLane~\cite{bergholm2018pennylane}. To ensure reproducibility and facilitate follow-up work, we extend torchdistill~\cite{torchdistill}. 
\subsection{Evaluation}
Our experiments are conducted on small-scale simulations with considerably fewer classes than the original work. Therefore, to draw meaningful insights from our results, the dimensions of the classical models are set equal to the number of qubits of their hybrid counterparts., i.e., the MLP first projects the high-dimensional backbone features to a low-dimensional classical embedding. For example, in an experiment with four qubits, we compare a hybrid predictor with a classical baseline predictor where the first layer of the MLP projects the backbone features to a four-dimensional representation.
\begin{table}[ht]
\centering
\caption{Training and Test Subsets of ILSVRC}
\label{tab:datasetsum}
\resizebox{\columnwidth}{!}{%
\begin{tabular}{@{}cccc@{}}
\toprule
Task      & Classes & Training Samples & Test Samples \\ \midrule
Nutriment & 10      & 13'000           & 500          \\
Felidae   & 13      & 16'900           & 650          \\
Buildings & 14      & 18'200            & 700          \\
Vessels   & 23      & 29'900           & 1150         \\ \bottomrule
\end{tabular}%
}
\end{table}
To emulate the scenario of \Cref{sec:disarch} with clients requesting inference from diverse environments, we create four thematically unrelated datasets from the 1000 labels from the ILSVRC classification task. \Cref{tab:datasetsum} summarizes each dataset representing a different location requiring separate predictors. The accompanying repository contains a script and instructions to recreate the datasets. \begin{table}[h]
\centering
\caption{Top-1 (Err)or of (C)lassic, (H)ybrid, Hybrid with (Res)iduals}
\label{tab:resultsum}
\resizebox{\columnwidth}{!}{%
\begin{tabular}{@{}ccccc@{}}
\toprule
                           & Qubits & Top-1 Err. C.  (\%) & Top-1 Err. H.  (\%) & Top-1 Err. H. Res. (\%) \\ \midrule
\multirow{4}{*}{Nutriment} & 4      & 13.00               & 37.13               & 12.11                   \\
                           & 6      & 11.73               & 25.20               & 10.40                   \\
                           & 8      & 11.30               & 16.90               & 10.07                   \\
                           & 10     & 10.69               & 14.80               & 9.58                    \\ \midrule
\multirow{4}{*}{Felidae}   & 4      & 19.82               & 31.57               & 19.05                   \\
                           & 6      & 17.60               & 29.07               & 16.77                   \\
                           & 8      & 16.77               & 18.77               & 15.85                   \\
                           & 10     & 16.56               & 18.31               & 15.31                   \\ \midrule
\multirow{4}{*}{Buildings} & 4      & 9.29                & 31.57               & 8.57                    \\
                           & 6      & 7.26                & 14.86               & 6.86                    \\
                           & 8      & 6.14                & 10.57               & 5.71                    \\
                           & 10     & 5.74                & 9.00                & 5.27                    \\ \midrule
\multirow{4}{*}{Vessels}   & 4      & 32.43               & 62.00               & 30.69                   \\
                           & 6      & 27.82               & 48.52               & 26.00                   \\
                           & 8      & 25.48               & 31.56               & 24.96                   \\
                           & 10     & 24.26              & 25.87               & 23.91                   \\ \bottomrule
\end{tabular}%
}
\end{table}

We run experiments with 4, 6, 8, and 10 qubits with a classical predictor as the baseline. The depth of the circuit is fixed at 8. We performed additional experiments with varying depth sizes and found that increasing the depth yields diminishing accuracy improvement.
\Cref{tab:resultsum} summarizes our results. The Plots in \Cref{plot:noresidualsresults} and \Cref{plot:residualsresults} show how a hybrid model without and with the skip connection compares to their classical counterpart for each dataset.
\begin{figure}[ht]
\centering
\includegraphics[width=\columnwidth]{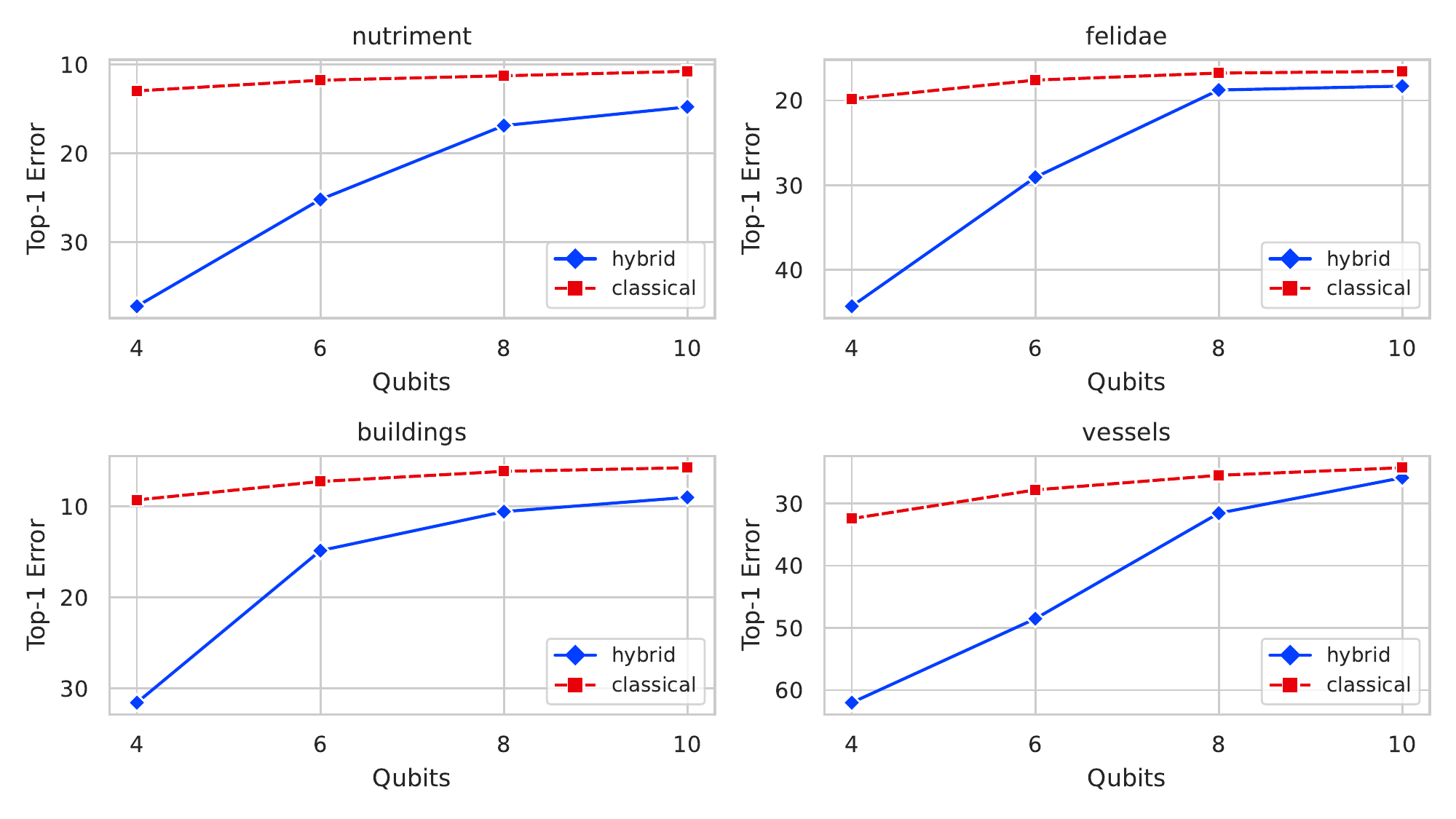}
\caption{Hybrid QNN without Skip Connection}
\label{plot:noresidualsresults}
\end{figure}
Relative to the classical baselines, Hybrid QNNs \textit{without} skip connections gradually approach comparable, albeit still worse, Top-1 error as we increase the number of qubits. For 2 and 4 qubits, the Top-1 error is roughly 20-30\% worse while the difference narrows to 2-5\%. Interestingly, Hybrid QNNs \textit{with} skip connections consistently outperform the classical baselines across all numbers of qubits. 
\begin{figure}[ht]
\centering
\includegraphics[width=\columnwidth]{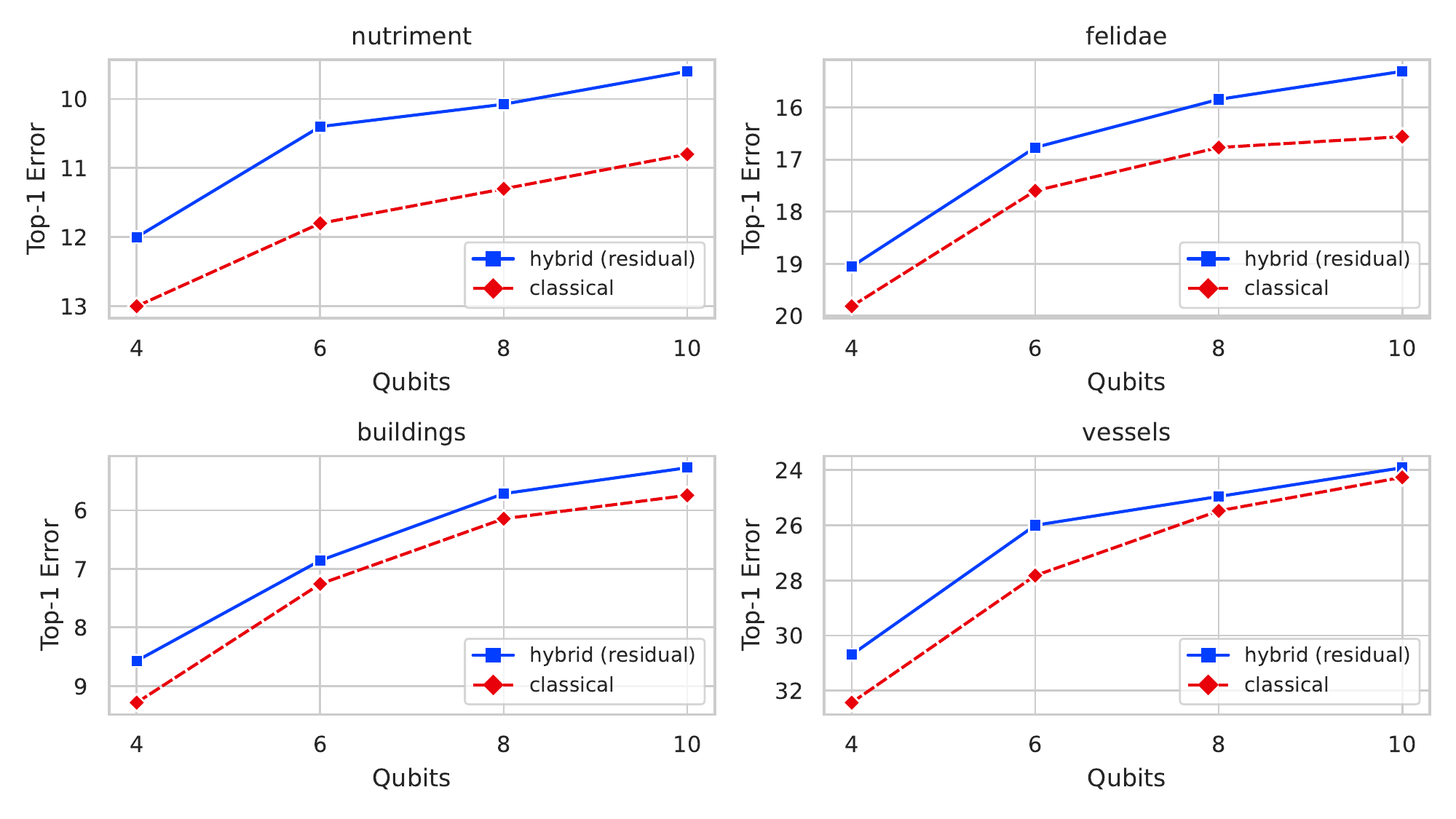}
\caption{Hybrid QNN with Skip Connection}
\label{plot:residualsresults}
\end{figure}
Although the motivation of skip connections in classical residual networks is to mitigate accuracy saturation for very deep neural networks, they essentially learn a residual function. Considering the autoencoder-backbone pair already heavily processes the input data, we hypothesized that a QNN could find more suitable representations for the classical features for some instances. In contrast, a QNN could decrease the performance for samples already sufficiently processed for classification. The QNN narrowing the performance gap with increasing qubits is consistent with our assumptions. The model without a skip connection cannot find a representation as good as the initial input for a low number of qubits. Conversely, the QNN with a skip connection can learn the residual function and sees a performance gain for the samples, where a quantum embedding is useful.  

A skip connection was the first intuitive attempt to provide empirical evidence, and the initial results seem promising. Nevertheless, we remind the reader that QuantenSplit only serves as a PoC to determine whether our vision is viable. The evaluation strategy is not extensive enough, and thus, our results should not be considered conclusive. Moreover, even with the skip connection, the hybrid model only marginally outperforms the classical model across all tasks simulations with a \textit{noise free} device. We did not experiment with optimization algorithms more appropriate for QNNs and did not spend considerable effort conceiving a suitable circuit design. Further, the backbone is classical, i.e., the extracted features may be biased favorably towards classical predictors. Future work can significantly improve our results by experimenting with more sophisticated approaches for mapping low-dimensional qubits to a high-dimensional feature space~\cite{peters2021machine}. Additionally, once training large QNN extractors is feasible, it would be interesting to determine whether we can train the classical compression model to find more suitable representations for quantum embeddings.

\section{Conclusion} \label{sec:conclusion}
This work presented our vision of integrating quantum computing into the edge-cloud continuum. We summarized existing literature in quantum and classical computing relevant to our work and subsequently described the importance of extending existing Classical Systems for the edge. Then, we introduced an architecture for a hybrid classical-quantum platform and identified the critical challenges of integrating QPUs. We focused on facilitating research efforts in quantum applications with warm-starting and AI inference for distributed intelligent tasks. Lastly, we extended a classical split inference method to support Hybrid QNNs optionally. Our results provide empirical evidence for the viability of our vision and suggest that our ideas are interesting research directives.

\section*{Acknowledgment}
We would like to thank
Fabian Bühler,
Benjamin Weder, Vladimir Yussupov and
Florian Kowarsch for their valuable input.
This work was partially funded by the BMWK projects \textit{PlanQK} (01MK20005N), \textit{EniQmA} (01MQ22007B), and \textit{SeQuenC} (01MQ22009B).


\bibliographystyle{plain}
\bibliography{main}
\end{document}